\documentclass[runningheads]{llncs}
\usepackage{graphicx}
\usepackage{subfigure}
\usepackage{amsmath}
\usepackage{color}
\usepackage{ulem}
\begin{document}
\title{Learning Geometry-Dependent and Physics-Based 
Inverse Image Reconstruction}
\titlerunning{Learning Geometry-Dependent Physics-Based Inverse Image Reconstruction}
%
\author{Xiajun Jiang \and
Sandesh Ghimire \and
Jwala Dhamala \and
Zhiyuan Li \and
Prashnna Kumar Gyawali \and
Linwei Wang}
%
\authorrunning{X. Jiang et al.}
%
\institute{Rochester Institute of Technology, Rochester, NY 14623, USA \\
\email{xj7056@rit.edu}}
\maketitle              

\begin{abstract}
Deep neural networks have shown great potential in image reconstruction problems in Euclidean space. However, many reconstruction problems involve imaging physics that are dependent on the underlying non-Euclidean geometry. In this paper, we present a new approach to learn inverse imaging that exploit the underlying geometry and physics. We first introduce a non-Euclidean encoding-decoding network that allows us to describe the unknown and measurement variables over their respective geometrical domains. We then learn the geometry-dependent physics in between the two domains by explicitly modeling it via a bipartite graph over the graphical embedding of the two geometry. We applied the presented network to reconstructing electrical activity on the heart surface from body-surface potential. In a series of generalization tasks with increasing difficulty, we demonstrated the improved ability of the presented network to generalize across geometrical changes underlying the data in comparison to its Euclidean alternatives.

\end{abstract}

\section{Introduction}

Deep learning has shown state-of-the-art performance in image reconstruction tasks across a variety of medical modalities \cite{zhu18,sun2016deep,Adler18,lucas18,ghimire2018generative}. These approaches typically formulate the problems in standard Euclidean image grids. However, in many problems, the unknown variables of interests and the corresponding measurements are defined over non-Euclidean geometrical domains: their physics-based relationship, both forward and inverse, is largely reliant on the underlying geometry. Examples include electrical activity in the heart and the potential it generates on the body surface \cite{bacoyannis2019deep,ghimire2019noninvasive,ghimire2017overcoming}, or electrical activity in the brain and its potential measurements on the skull surface \cite{michel2012towards}. Standard Euclidean deep learning neglecting the underlying geometry not only ignores the geometry-dependent imaging physics, but also has difficulty in generalizing over different geometry.

To design inverse imaging (image reconstruction) networks that can generalize across geometry, there are two general approaches. One is to make the network invariant to geometry by, for instance, an information bottleneck that removes geometrical information from the input data \cite{ghimire2019improving}. While demonstrating improved generalization to geometrical changes \cite{ghimire2019improving}, the treatment of non-Euclidean data as Euclidean data ties the network to the training mesh and prevents its direct application to unseen meshes from new patients. Alternatively, one can make the network equivariant to the geometry. In \cite{bacoyannis2019deep}, for instance, the reconstruction of electrical activity in the heart is formulated and conditioned on 2D image scans of the heart \cite{bacoyannis2019deep}.   Rather than explicitly describing the geometry, this approach defines non-Euclidean variables at a small region of interest within the Euclidean image grid. How to extend it to consider the geometry of both the unknown (\textit{e.g.}, the heart) and the measurement (\textit{e.g.}, the body), and to explicitly consider the geometry-dependent physics in between, is not clear.

Graph convolutional neural networks (GCNN) provide an appealing alternative to solving inverse imaging between non-Euclidean variables defined over geometrical domains \cite{bronstein2017geometric}. Significant efforts in GCNN have been made for node- and graph-level classifications,  graph embedding, and graph generation \cite{wu2020comprehensive}. However, no existing work has considered learning geometry-dependent relationship between signals defined on two separate graphs, which is a critical component of achieving physics-based inverse imaging. 

In this paper, we present a non-Euclidean inverse imaging (image reconstruction) network that 1) directly models the unknown and its measurement over their geometrical domains, and 2) models and learns their inverse relationship -- as informed by the physics -- as a function of the geometry. It consists of two novel contributions. First, to describe the spatiotemporal variables (unknowns and measurements) over their respective geometrical domain, we introduce an encoding-decoding architecture composed of spatial-temporal graph convolutional neural networks (ST-GCNN) defined separately for each domain. Second, to learn the geometry-dependent physics in between, we model it with a bipartite graph between the graphical embedding of the two geometrical domains. We applied the presented method for reconstructing spatiotemporal electrical potential on the ventricular surface from body-surface potential. In synthetic and real-data experiments, we tested the presented network in a series of generalization tasks with increasing difficulty, and compared it to Euclidean baselines without and with a geometry-invariant bottleneck \cite{ghimire2019improving}. By learning inverse imaging in a geometry-dependent and physics-informed fashion, the presented network showed an improved generalization to geometrical changes in the data.

\section{Methodology}

Cardiac electrical excitation produces time-varying voltage signals on the body surface, following quasi-static approximation of the electromagnetism \cite{plonsey2001bioelectric}. Given a pair of heart and torso geometry, the governing physics can be numerically approximated to relate signals in the heart $\mathbf{X}_t$ to those on the body surface $\mathbf{Y}_t$:
\begin{equation}
    \mathbf{Y}_t
    = \mathbf{H} \mathbf{X}_t
    \quad \forall t \in \{1,...,T\}.
    \label{bemrforward}
\end{equation}

Note that $\mathbf{X}_t$ and $\mathbf{Y}_t$ live on the 3D geometry of the heart and torso surface, respectively. The forward operator $\mathbf{H}$ defines the physics of their relationship and is highly dependent on the given heart-torso geometry. Traditional approaches to reconstructing $\mathbf{X}_t$ from $\mathbf{Y}_t$ starts with this forward model, exploiting the geometry and physics behind the inverse relationship. When using Euclidean deep learning for direct inference of $\mathbf{X}_t$ from $\mathbf{Y}_t$, the network becomes solely reliant on labeled data pairs, incorporating neither the physics nor the geometry underlying the problem. The proposed method is set to bridge these gaps by 1) allowing the description of $\mathbf{X}_t$ and $\mathbf{Y}_t$ in their geometrical domains, and 2) explicitly modeling their physics relationship as a function of the geometry.

\begin{figure}[t]
\centering
\includegraphics[width=\textwidth]{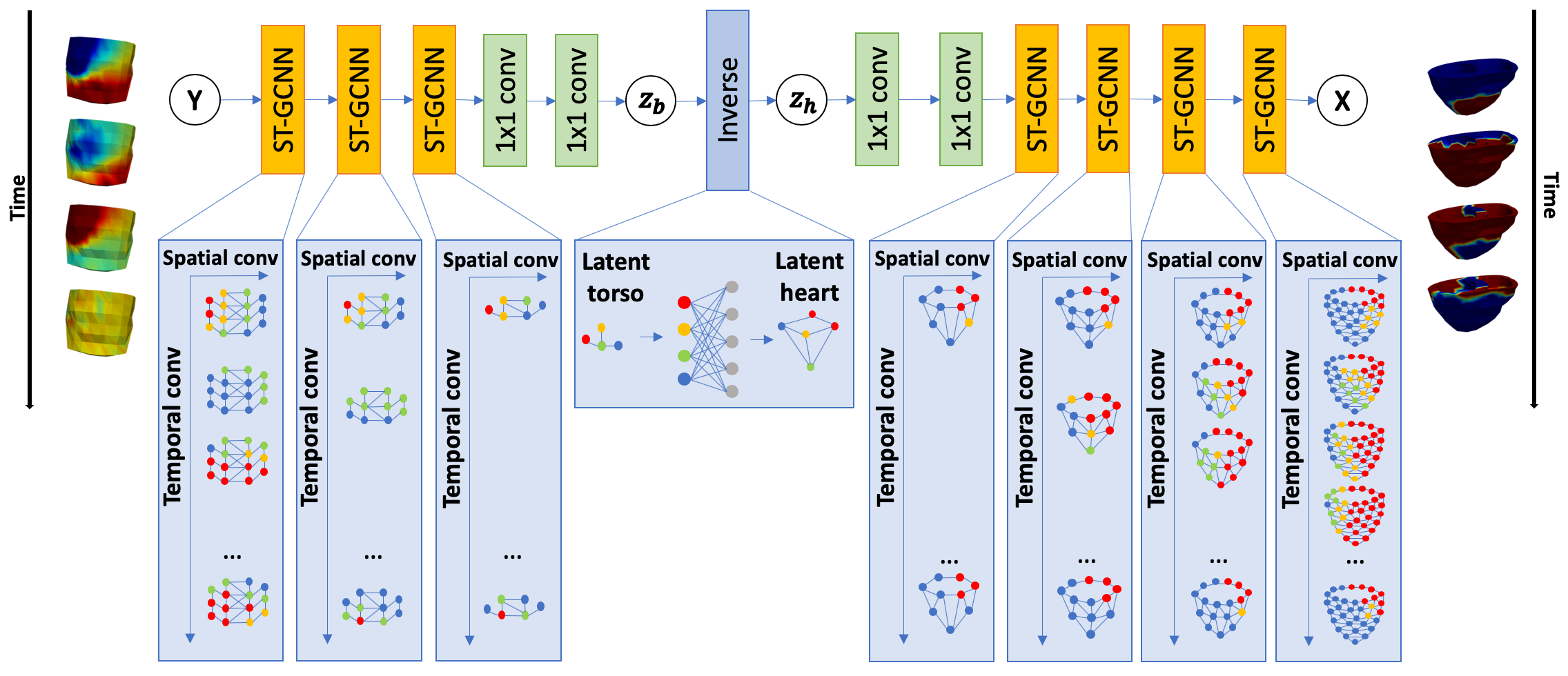}
\caption{Illustration of the presented non-Euclidean inverse imaging network.
} 
\label{fig:overview}
\end{figure}

As summarized in Fig.~\ref{fig:overview}, we present an encoder-decoder architecture with ST-GCNNs to embed/generate $\mathbf{Y}_t$ and $\mathbf{X}_t$ over their respective geometry. The geometry-dependent relationship between the latent variables of $\mathbf{Y}_t$ and $\mathbf{X}_t$ is learned via a bipartite graph over the graph embedding of the two geometry.

\subsection{Encoding-Decoding with ST-GCNNs} 

As $\mathbf{X}_t$ and $\mathbf{Y}_t$ are temporal sequences living on 3D geometry, we describe their generation/embedding with ST-GCNNs that consist of interlaced graph convolution in space and regular convolution in time. As illustrated in Fig.~\ref{fig:overview}, both spatial and temporal dimensions are reduced/expanded during encoding/decoding.

\subsubsection{Geometrical Representation in Graphs:}
We represent triangular meshes of the heart and torso as two separate undirected graphs: $\mathcal{G} = (\mathcal{V}, \mathcal{E}, \mathbf{U}, \mathbf{F})$, where vertices $\mathcal{V}$ consist of all $V$ mesh nodes and edges $\mathcal{E}$ describe the vertex connection as defined by the triangular mesh. $\mathbf{U} \in [0, 1]^{V \times V \times 3}$ consists of edge attributes $\mathbf{u}(i, j)$ between vertex $i$ and $j$ as normalized differences in their 3D coordinates $((x_i - x_j) / s, (y_i - y_j) / s, (z_i - z_j) / s)$ if an edge exists, and 0 otherwise, where $s = \sqrt{(x_i - x_j)^2 + (y_i - y_j)^2 + (z_i - z_j)^2}$. $\mathbf{F} \in \mathrm{R}^{V \times M \times T}$ represents the time sequences of node features across all vertices.

\subsubsection{Spatial Graph Convolution:}
A continuous spline kernel for spatial convolution is used such that it can be applied across graphs \cite{fey2018splinecnn}. Given graph node features $\mathbf{f} \in \mathrm{R}^{V \times M}$ at each time instant, the convolution kernel is defined as: 
\begin{equation}
    g_l(\mathbf{u}) = \sum_{\mathbf{p} \in \mathcal{P}} w_{\mathbf{p}, l} B_{\mathbf{p}}(\mathbf{u}),
    \label{spline-kernel}
\end{equation}
where $1 \leq l \leq M$, the spline basis  $B_{\mathbf{p}}(\mathbf{u}) = \prod_{r=1}^d N_{r, p_r}^m(\mathbf{u})$ with $N_{r, p_r}^m$ denoting $d$ open B-spline basis of degree $m$ based on equidistant knot vectors, $\mathcal{P} = (N_{1, r}^m)_r \times ... \times (N_{d, r}^m)_r$ is the Cartesian product of the B-spline bases, and $w_{\mathbf{p}, l}$ are trainable parameters. Given kernel $\mathbf{g} = (g_1,...,g_M)$, spatial convolution for vertex $i\in \mathcal{V}$ with its neighborhood $N(i)$ is defined as
\begin{equation}
    (f_l * g_l)(i) = \sum_{j \in N(i), \mathbf{p} \in \mathcal{P}(\mathbf{u}(i, j))} f_l(j) \cdot g_l(\mathbf{u}(i, j)).
    \label{spline-cnn}
\end{equation}
Since the B-spline basis in equation (\ref{spline-kernel}) is conditioned on local geometry, the learned kernel can be applied across graphs and the convolution incorporates geometrical information within the graph. This spatial convolution is independently applied to each time frame of the signal sequence in parallel.

To make the network deeper and more expressive, we introduce residual blocks here to pass the input of spatial convolution through a skip connection with 1D convolution before adding it to the output of the spatial convolution. 

\subsubsection{Temporal Modeling:} 
After spatial convolution, temporal convolution using standard 1D convolution is applied to the time sequence for each node and feature. The number of filters is set to compresses the time sequence in dimension in the encoder, while expanding in the decoder.
The geometry graph remains the same for the complete temporal sequences.

\subsubsection{Hierarchical Graph Composition:}
To allow pooling and unpooling in space, we further introduce a hierarchical graph representation of the two geometry. While various graph clustering \cite{dhillon2007weighted} and pooling methods \cite{wu2020comprehensive} exist, a unique constraint needs to be met here due to the underlying physics: the topology of the geometry must be preserved in its hierarchical representations to prevent non-physical spatial propagation of signals. Here, we obtain hierarchical geometry representations by specialized mesh coarsening method in CGAL \cite{cgal:c-tsms-12-20a,lindstrom1998fast}.

The hierarchical graph representation is predefined and stored in matrices to allow efficient matrix multiplications for pooling/unpooling \cite{dhamala2019bayesian}. If $\mathcal{G}_o$ is a graph with $N_1$ vertices and $\mathcal{G}_c$ is its coarsened graph with $N_2$ vertices, we use a binary matrix $\mathbf{P} \in \mathrm{R}^{N_1 \times N_2}$, where $\mathbf{P}_{ij} = 1$ if vertex $i$ in $\mathcal{G}_o$ is grouped to vertex $j$ in $\mathcal{G}_c$, and $\mathbf{P}_{ij} = 0$ otherwise. Given feature map $\mathbf{f}_o \in \mathrm{R}^{N_1 \times M}$ over $\mathcal{G}_o$ and $\mathbf{f}_c \in \mathrm{R}^{N_2 \times M}$ over $\mathcal{G}_c$, the pooling operation is defined by $\mathbf{f}_c = \mathbf{P}_n^T \mathbf{f}_o$ and the unpooling operation is defined by $\mathbf{f}_o = \mathbf{P} \mathbf{f}_c$, where $\mathbf{P}_n^T$ is column normalized from $\mathbf{P}$.

\subsubsection{Summary:}  
As summarized in Fig.~\ref{fig:overview}, each ST-GCNN block consists of spatial graph convolution, temporal convolution, and spatial pooling/unpooling as described above. Using these building blocks, we obtain an encoder that embeds body-surface signal $\mathbf{Y}_t$ over its torso geometry, and a decoder that generates heart-surface potential $\mathbf{X}_t$ over its heart geometry. Next, we learn the physics-based relationship between the two latent space as a function of their geometry.

\subsection{Learning Geometry-Dependent Physics in Latent Space}
As explained earlier, the physics between $\mathbf{X}_t$ and $\mathbf{Y}_t$ is heavily reliant on the underlying heart-torso geometry: according to equation (\ref{bemrforward}), the potential on one torso node can be represented as a linear combination of the potential from all heart nodes, where the coefficients are determined by the relative position between each pair of torso-heart nodes. We assume the linearity to hold between the heart and torso signals in the latent space during inverse imaging, and explicitly model it as a function of the relative position between embedded heart and torso geometry, where a quadratic function exists between the coefficients of the linear function and the geometry.

To do so, we construct a bipartite graph where the edge exists between each pair of heart and torso vertices from their respective graph embedding: the edge attribute $\mathbf{u}(i, j)$ between torso vertex $i$ and heart vertex $j$ thus describes their relative geometrical relationship. The bipartite graph is also learned using the complete temporal sequences. For latent representation $\mathbf{z}_h(i)$ on vertex $i$ of the latent heart mesh, we define it as a linear combination of latent representation $\mathbf{z}_b(j)$ across all vertices $j$ of the latent torso mesh: 
\begin{equation}
    \mathbf{z}_h(i) = \sum_{j} \mathbf{z}_b(j) \cdot \hat{\mathbf{h}}(\mathbf{u}(i, j)),
    \label{invele}
\end{equation}
where the coefficients $\hat{\mathbf{h}}(\mathbf{u}(i, j))$ are dependant on the relative position $\mathbf{u}(i, j)$ between the two graphs. Aside from being a physics-informed function, this geometric parameterization allows the learned function to generalize across different torso-heart geometry. None of this would be achievable by, for instance, using fully connected layers between $\mathbf{z}_b$ and $\mathbf{z}_h$. Exploiting the similarity between eq.(\ref{invele}) and eq.(\ref{spline-cnn}), we recast linear relationship in eq.(\ref{invele}) using spline convolution, with the geometry-dependent coefficients $\hat{\mathbf{h}}$ learned as the spline convolution kernel.  

\subsection{Loss Function}
Denoting the encoder as $\mathbf{z}_b = E_\theta(\mathbf{Y})$, the geometry-dependent inverse function as $\mathbf{z}_h = h_\rho(\mathbf{z}_b)$, and the decoder as as $\mathbf{X} = D_\phi(\mathbf{z}_h)$, parameters $\theta, \rho$ and $\phi$ of the network are optimized by minimizing the mean square error between the reconstructed $\hat{\mathbf{X}}_i$ on given pairs of training data $\{ \mathbf{X}_i, \mathbf{Y}_i \}_{i=1}^N$: 
\begin{equation}
    \mathcal{L} =\sum_i || \mathbf{X}_i -  D_\phi \left(h_\rho \left(E_\theta \left(\mathbf{Y}_i \right) \right) \right) ||_2^2.
\end{equation}

\section{Experiments} 
We design a series of generalization tasks of increasing difficulty. In specific, we trained the network using synthetic data simulated on a specific pair of heart-torso geometry, including geometrical variations introduced by rotating the heart along the longitudinal axis (z-axis) for a predefined range. We then tested the trained network regarding generalization to: 1) synthetic data simulated on the same heart-torso geometry but with z-axis heart rotations beyond the training range, 2) synthetic data simulated on the same heart-torso geometry but with novel heart rotations along frontal axis (x-axis) and sagittal axis (y-axis), 3)  synthetic data simulated on new heart-torso geometry from new patients, and 4) real data on different heart-torso geometry.

The first two tests considered comparisons to Euclidean encoding-decoding networks \cite{ghimire2019improving}, both in a deterministic formulation and in a stochastic formulation with improved invariance to input geometry. These Euclidean networks will not apply without re-training on the new geometry in the last two tests.

\subsubsection{Models and Training:}
In all experiments, the presented network consists of three ST-GCNN blocks and two standard convolutional layers in the encoder, one spline convolutional layer in the inverse block, and four ST-GCNN blocks and two standard convolutional layers in the decoder. We used ELU activation, ADAM optimizer \cite{kingma2014adam}, and a learning rate of $5 \times 10^{-4}$. The Euclidean baselines followed the architectures presented in \cite{ghimire2019improving}, which consist of cascaded LSTMs and fully connected layers in the encoder and decoder. 

For training, we generated pairs of simulated potential data on a specific heart-torso mesh. On the heart, we simulated spatiotemporal propagation sequence of action potential by the Aliev-Panfilov (AP) model \cite{aliev1996simple}, considering a combination of 38 different origins of activation and 16 spatial distribution of scar tissue in the heart. We then rotated the heart by -2$^\circ$ to 2$^\circ$ around the z-axis, obtaining approximately 2700 sets of different body-surface potential embodying changes of heart orientations in the data. All body-surface potential were corrupted with 20dB Gaussian noises for inverse imaging. Using NVIDIA Tesla T4 with 16 GB memory, the geometric model took 3 days for training.

Synthetic data for testing were generated in a similar fashion, with additional geometry changes as detailed in later sections. The reconstruction accuracy was measured by the mean square  error (MSE) and correlation  coefficient (CC)  between  the  reconstructed  and  actual  potential sequence on the heart surface.

\begin{figure}[t]
    \centering
    \includegraphics[width=\textwidth]{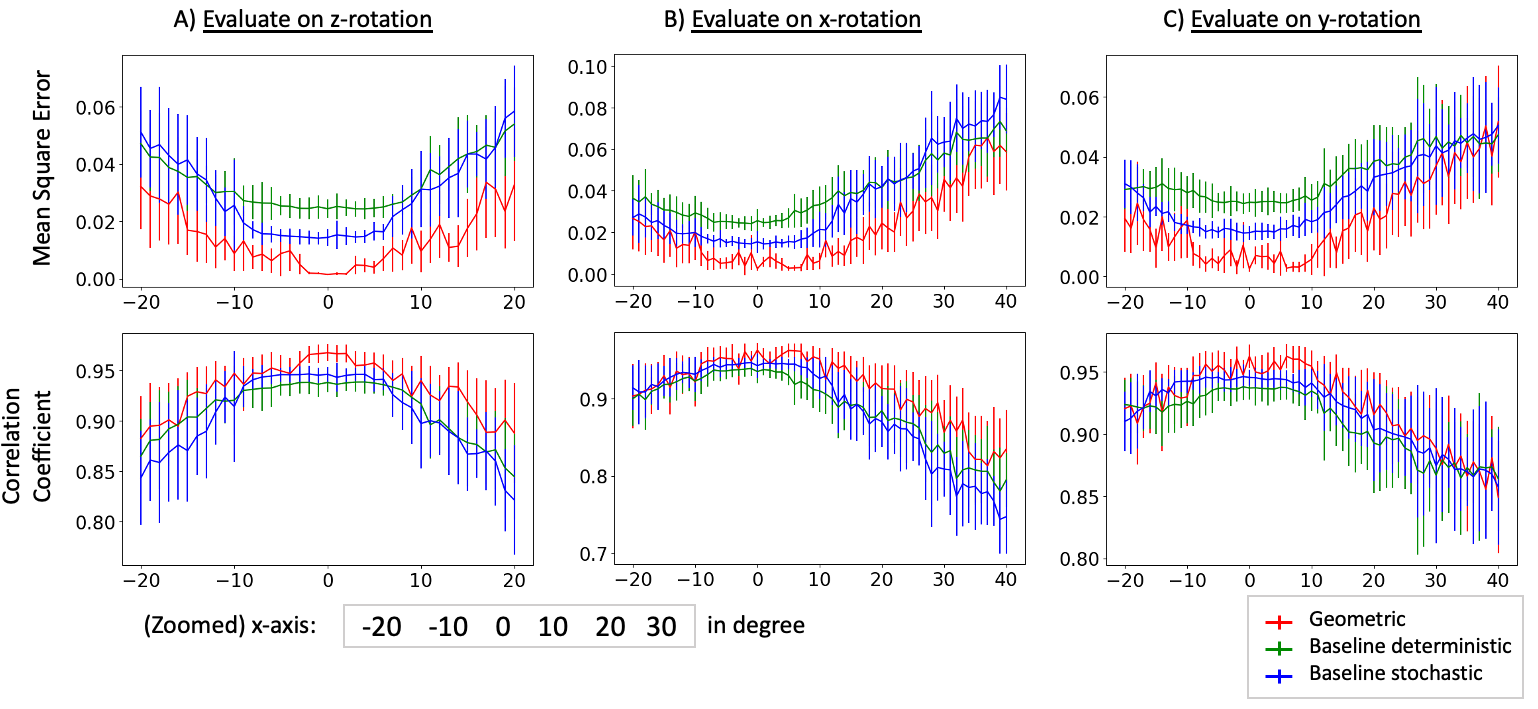}
    \caption{Comparison of reconstruction accuracy 
    among the three comparison models in test data with A) heart rotations outside training range, and B)-C) novel rotations not seen in training. X-axis represents the degree of rotation relative to training.}
    \label{fig:exp1}
\end{figure}

\begin{figure}[!ht]
    \centering
    \includegraphics[width=.8\textwidth]{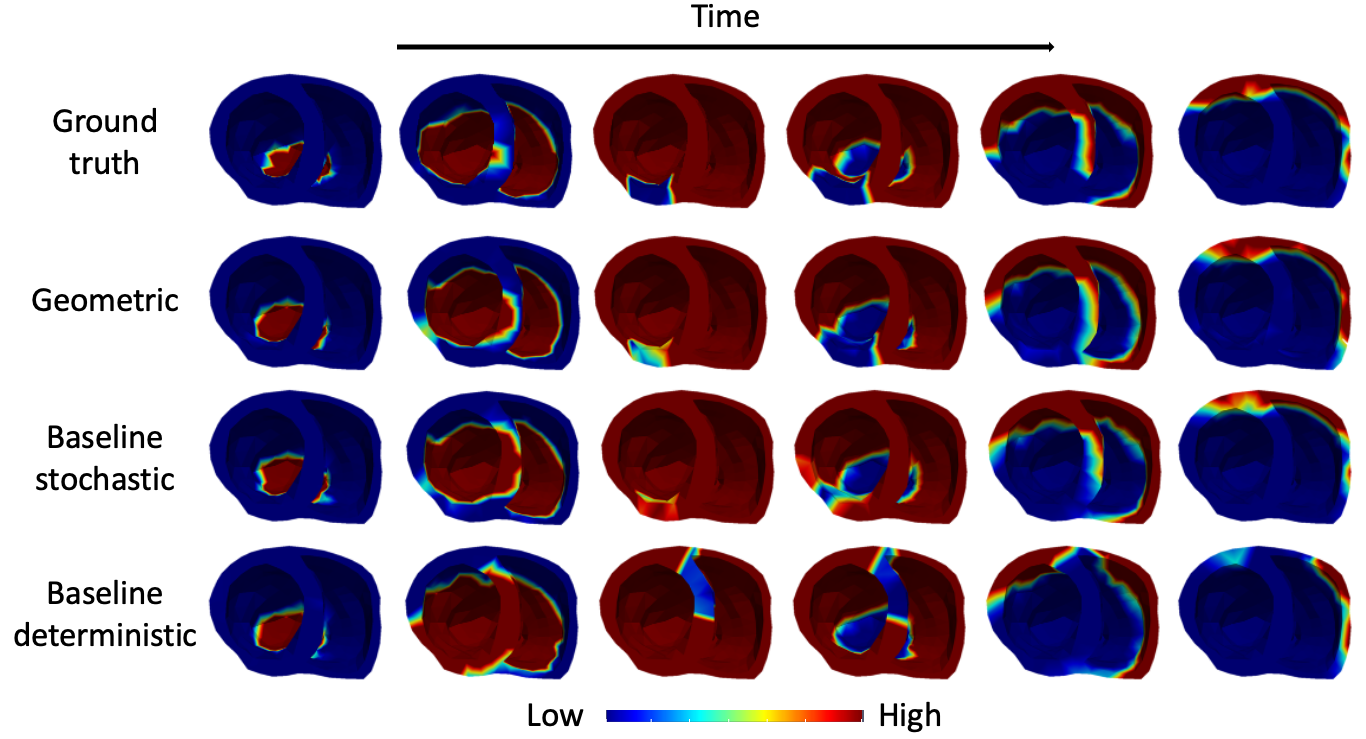}
    \caption{Reconstructed electrical activity by three comparison models when z = -19$^\circ$.}
    \label{fig:exp11_view}
\end{figure}

\subsubsection{Generalization to Rotations Outside Training Range:}
We first applied the trained models to body-surface potential data generated when the heart was rotated by -20$^\circ$ to 20$^\circ$ around the z-axis, a range far outside that considered in training. Fig.~\ref{fig:exp1}A summarizes the quantitative metrics of the three models on approximately 22,000 test cases, against the change in heart rotations from training data. As shown, the presented method (red) outperformed the deterministic (green) and stochastic (blue) Euclidean baseline  in all metrics for all heart rotations. The standard deviation of the geometric method lies in between that of deterministic and stochastic baseline.

\subsubsection{Generalization to Novel Rotations:}
We then tested the trained models on approximately 66,000 body-surface data generated from novel heart rotations around the x-axis (-20$^\circ$ to +40$^\circ$) and y-axis (-20$^\circ$ to +40$^\circ$). As summarized in Fig.~\ref{fig:exp1}B-C, the presented model (red) significantly outperformed the two Euclidean models in all metrics. Furthermore,  we observe in Fig.~\ref{fig:exp1}B that the geometric method performs better as the test set deviates more from the training set (up to 40 degree of rotation). This supports that, as test data move further away from training, the gain in the generalization ability of the presented method would become more significant in comparison to its Euclidean alternatives. The standard deviations of the three models are comparable.

\subsubsection{Generalization to New Geometry:}
\begin{figure}[t]
    \centering
    \includegraphics[width=.8\textwidth]{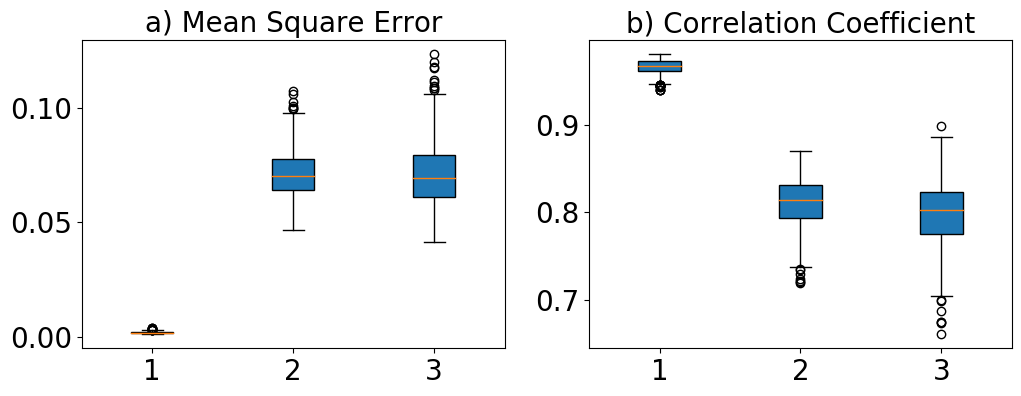}
    \caption{Accuracy of reconstruction on training geometry (1) and new geometry (2/3).}
    \label{fig:exp2}
\end{figure}

We then moved to apply the trained network to simulated data generated on two new heart-torso meshes. This represents a realistic scenario where the network trained on a group of patients will be applied to new patients. Fig.~\ref{fig:exp2} provides box plots of the two metrics obtained on the two new geometry over, respectively, 491 and 444 test data. Despite a drop in performance in comparison to the earlier results on the training geometry, reasonable accuracy was achieved considering the difficulty of the generalization task. Note that Euclidean networks will not be applicable here unless being re-trained on data generated on the new geometry \cite{ghimire2019improving}. 

\subsubsection{Generalization to Real Data:}
\begin{figure}[t]
    \centering
    \includegraphics[width=.9\textwidth]{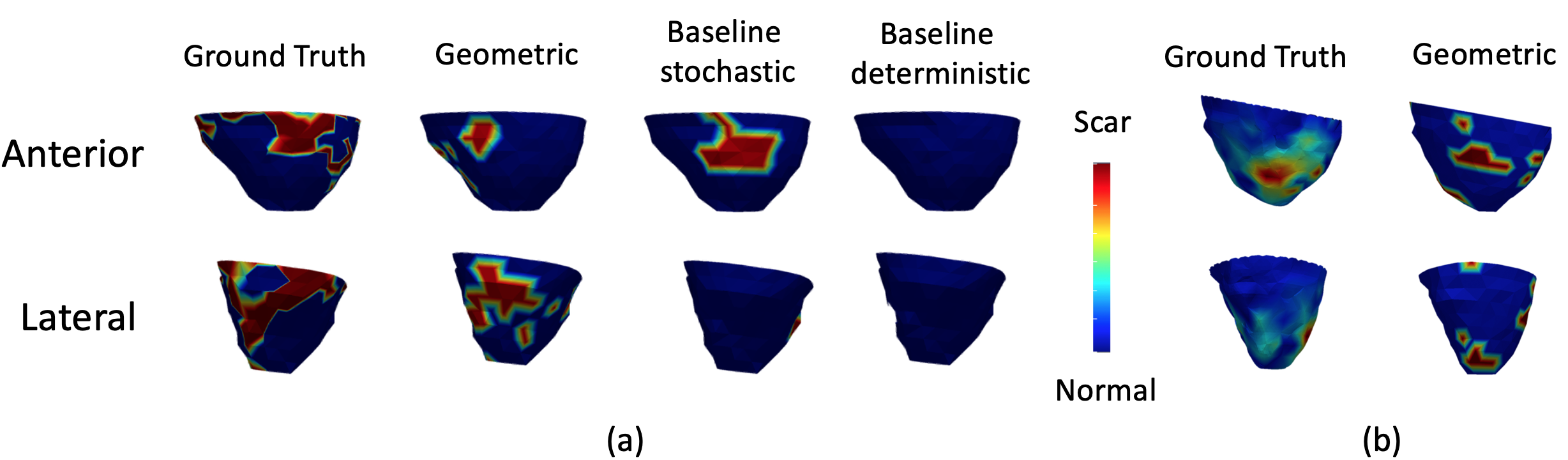}
    \caption{Region of scar identified from reconstructed potential sequence on a) the training patient and b) a new patient.
    The ground truth was from \textit{in-vivo} voltage mapping.}
    \label{fig:exp3}
\end{figure}

Finally, we tested the presented network on \textit{in-vivo} 120-lead body-surface potential data obtained on two patients with scar-related ventricular arrhythmia. Since the heart-torso geometry of patient $\sharp 1$ was used in training, we were able to apply the Euclidean baselines for comparison purpose. From each reconstructed potential sequence on each patient, we identified the region of scar tissue by nodes whose activation was shorter than a predefined duration. The results summarized in Fig.~\ref{fig:exp3} demonstrated the ability of the presented network to not only generalize across geometry but across the shifts between simulated and real data, approximating the location of scar tissue with evident visual improvement over its Euclidean alternatives. 

\section{Conclusion}

In this work, we present a novel non-Euclidean network for learning geometry-dependent and physics-based inverse imaging between spatiotemporal variables living on 3D geometrical domains. In generalization tests with increased difficulty, we demonstrated the ability of the presented network to better generalize to unseen geometrical variations in comparison to its Euclidean alternatives, and to directly apply to new geometry which is not possible with Euclidean approaches. An immediate future work is to explore the use of fine-tuning with a small number of labeled data in order to improve the performance of the network when applying it to new patients. To our knowledge, this is the first geometry-dependent inverse imaging network over non-Euclidean domains and its application to reconstructing cardiac electrical activity from surface potential.

%
%
%
\bibliographystyle{splncs04}
\bibliography{paper2300}

\end{document}